\documentclass[pre,aps,preprint,amsmath]{revtex4}

\usepackage{graphicx}
\usepackage{dcolumn}
\usepackage{bm}
\usepackage{latexsym,bm,amsmath,amssymb}

\begin{document}

\title{Transformation thermal convection: Cloaking, concentrating, and camouflage}


\author{Gaole Dai}
\author{Jin Shang}
\author{Jiping Huang}\email{jphuang@fudan.edu.cn}

\affiliation{Department of Physics, State Key Laboratory of Surface Physics, and Key Laboratory of Micro and Nano Photonic Structures (MOE), Fudan University, Shanghai 200433, China}
\affiliation{Collaborative Innovation Center of Advanced Microstructures, Nanjing 210093, China}

\date{\today}

\begin{abstract}


Heat can generally transfer via thermal conduction, thermal radiation, and thermal convection. All the existing theories of transformation thermotics and optics can treat thermal conduction and thermal radiation, respectively. Unfortunately, thermal convection has never been touched in transformation theories due to the lack of a suitable theory, thus limiting applications associated with heat transfer through fluids (liquid or gas). Here, we develop, for the first time, a general theory of transformation thermal convection by considering the convection-diffusion equation, the Navier-Stokes equation, and the Darcy law. By introducing porous media, we get a set of coupled equations keeping their forms under coordinate transformation.  As model applications, the theory helps to show the effects of cloaking, concentrating, and camouflage. Our finite element simulations confirm the theoretical findings. This work offers a general transformation theory for thermal convection, thus revealing some novel behaviors of thermal convection; it not only   provides new hints on how to control heat transfer by combining thermal conduction, thermal radiation, and thermal convection, but also benefits the study of mass diffusion and other related fields that contain a set of equations and need to transform velocities at the same time.

\end{abstract}




\maketitle


With the advent of the energy crisis, non-renewable energy resources are decreasing while they are producing more and more waste heat. Thus, an efficient control of heat transfer becomes particularly crucial. It is well known that the heat can generally transfer via three mechanisms, namely, thermal conduction, thermal radiation, and thermal convection.

Regarding the thermal conduction, it has been extensively studied by developing the theories of transformation thermotics since 2008~\cite{APL2008,ChenAPL08,PRL2012,PRL2013,XuPRL14,HanPRL14,MaPRL14,Qiu,PRL2015,PRL2016}. As a result, some novel thermal metamaterials like thermal cloaks~\cite{APL2008,ChenAPL08,PRL2012,PRL2013,XuPRL14,HanPRL14,MaPRL14,Qiu,PRL2015,PRL2016}, thermal concentrators~\cite{PRL2012}, thermal rotators~\cite{PRL2012}, thermal camouflage~\cite{Qiu}, and macroscopic thermal diodes~\cite{PRL2015} have been theoretically designed and/or experimentally realized, which pave a new way to control the conduction of heat.

Regarding the thermal radiation, the existing theories of transformation optics~\cite{S2006,LeonhardtS06} can be adopted to handle it directly because thermal radiation is essentially associated with radiated electromagnetic waves.

However, the thermal convection has never been touched in transformation theories, even though transformation theories (say,  transformation thermotics and transformation optics as mentioned above) have been shown to be powerful for revealing novel behaviors like cloaking, concentrating, or camouflage. This is because of the lack of a suitable theory for transforming thermal convection.  The difficulty might be that one should consider a set of complex equations at the same time since the flow of fluid can influence heat transfer, while in thermal conduction, one only needs to transform the Fourier law. This fact largely limits applications associated with heat transfer through fluids (liquid or gas), which includes the process of thermal convection.

Actually, the theory of transformation thermotics~\cite{APL2008} is a thermal-conduction analogy of transformation optics~\cite{S2006}; the latter is based on the fact that the Maxwell equations can keep their forms under coordinate transformation. In principle, if we have a set of equations with form invariance in different coordinate systems, we can also get a self-consistent theoretical framework to control thermal convection (corresponding to heat transfer with flow movement simultaneously). However, the challenge is that the equations for electric field and magnetic field have the same form while equations considered in this work don't. To this end, we manage to establish a general theory of transformation thermal convection by considering the convection-diffusion equation, the Navier-Stokes equation, and the Darcy law.
 By introducing porous media, we successfully get a set of coupled equations keeping their forms under coordinate transformation.  As a model application, the theory helps to show the effects of cloaking, concentrating, and camouflage. Our finite element simulations confirm the theoretical findings.


{\it Theory of transformation thermal convection.}

For treating heat transfer in fluids, we start by modifying the Fourier law of heat conduction for incompressible flow without heat sources (neglecting the viscous dissipation term)~\cite{Landau} as
\begin{equation}\label{heat transfer}
\rho C_{p}\frac{\partial T}{\partial t}+\rho C_{p}\nabla\cdot (\vec{v}T)=\nabla \cdot(\kappa\nabla T),
\end{equation}
where $\rho$, $C_{p}$, and $\kappa$ are respectively the density, specific heat at constant pressure, and thermal conductivity of fluid materials. $\partial T/\partial t$ denotes the derivative of temperature $T$ with respect to time $t$, and $\vec{v}$ is the velocity of fluids. As is known, $ \rho C_{p}\nabla\cdot (\vec{v}T) $ is the term due to thermal convection.



For the coordinate transformation $\{x_{i}\}\rightarrow\{y_{j}\}$ and the associated Jacobian Matrix $\textbf{J}=\frac{\partial(y_{1},y_{2},y_{3})}{\partial(x_{1},x_{2},x_{3})},$ one can write
\begin{equation}
\frac{\rho C_{p}}{\det\textbf{J}}\left [\frac{\partial T}{\partial t}+\sum\limits_{j}\frac{\partial }{\partial y_{i}} \left (\sum\limits_{i}J_{ij}^{\intercal}v_{i}T\right )\right ]=\sum\limits_{ijkl}\frac{\partial}{\partial y_{k}}\left (\frac{1}{\det\textbf{J}}J_{ki}\kappa_{ij}J_{jl}^{\intercal}\frac{\partial T}{\partial y_{l}}\right ) .
\end{equation}
Let $ \rho' C_{p}'=\frac{\rho C_{p}}{\det\textbf{J}}, \vec{v}'=\textbf{J}^{\intercal}\vec{v} $
and $ \kappa'=\frac{\textbf{J}{\kappa}\textbf{J}^{\intercal}}{\det\textbf{J}}, $ and then we achieve
\begin{equation}\label{heat transfer transformed}
\rho' C_{p}'\left [\frac{\partial T}{\partial t}+\nabla'\cdot(\vec{v}'T)\right ]=\nabla'\cdot(\kappa'\nabla'T).
\end{equation}
We can also choose to write $ \vec{v}'=\frac{\textbf{J}^{\intercal}\vec{v}}{\det\textbf{J}} $ and don't change $ \rho $ or $ c_{p} $ for steady states. Eqs.~$\left(\ref{heat transfer}\right)$~and~$\left(\ref{heat transfer transformed}\right)$ have the same form, so we can develop transformation thermotics including heat convection.

Now, the key problem is how to realize both velocity distribution $\vec{v}'(\vec{r},t)$ and anisotropic thermal conductivity $ \kappa' $ of fluid materials. Generally, in order to describe the state of fluids completely, we need to know the velocity $ \vec{v} $ and any two thermodynamic quantities like $ \rho $ and pressure $ p $, which are determined by Eq.~$\left(\ref{heat transfer}\right)$ together with Navier-Stokes equations and the equation of continuity~\cite{Landau}
\begin{equation}\label{navier stokes}
(\vec{v}\cdot\nabla)\vec{v}=-\frac{1}{\rho}\nabla p+\frac{\eta}{\rho}\nabla\cdot\nabla\vec{v},
\end{equation}	
\begin{equation}\label{conservation law}
\nabla\cdot\vec{v}=0.
\end{equation}
For simplicity, we have assumed $ \vec{v}(\vec{r},t)=\vec{v}(\vec{r}) $ and $ \rho(\vec{r},t)\equiv\rho $. However, Eq.~$\left(\ref{conservation law}\right)$ can keep its form under coordinate transformation while Eq.~$\left(\ref{navier stokes}\right)$ doesn't in most cases, even though we can neglect nonlinear term $ (\vec{v}\cdot\nabla)\vec{v} $ when Reynolds number Re is small (a similar case as the elastic equation in~\cite{NJP2006}). It also remains a difficulty in experiments to make $ \kappa' $ anisotropic in fluids while this has been done successfully in heat conduction of solid materials. Fortunately, recent advances on velocity control~\cite{PRL2011} inspires us to consider heat transfer and velocity control in porous media at the same time.

In saturated porous media, we have a set of equations as~\cite{Bear1,Bear2}
\begin{equation}\label{heat transfer porous}
(\rho C_{p})_{m}\frac{\partial T}{\partial t}+\rho_{f} C_{p,f}(\vec{v}\cdot \nabla T)=\nabla \cdot(\kappa_{m} \nabla T),
\end{equation}
\begin{equation}\label{darcy's law}
\nabla p+\frac{\eta}{k}\vec{v}=0,
\end{equation}
\begin{equation}\label{conservation law 2}
\nabla\cdot\vec{v}=0,
\end{equation}
where $ \eta $ is the dynamic viscosity, $ \phi $ is the porosity and $ k $ is the permeability of porous media. The parameters with subscript $m$  denote those of porous media. In addition, $ \rho_{f}, C_{p,f} $ and $\kappa_{f} $ denote respectively the density, specific heat at constant pressure and thermal conductivity of fluid materials.  By taking volume-averaging method~\cite{Bear2}, the effective product of density and specific heat is defined as
\begin{equation}
(\rho C_{p})_{m}=(1-\phi)(\rho_{s} C_{p,s})+\phi(\rho_{f} C_{p,f})
\end{equation}
where $ \rho_{s}, C_{p,s} $ and $\kappa_{s} $ are respectively the density, specific heat at constant pressure and thermal conductivity of solid skeleton in porous media. And the effective conductivity is
\begin{equation}
\kappa_{m}=(1-\phi)\kappa_{s}+\phi\kappa_{f}.
\end{equation}
In Eq.~$\left(\ref{heat transfer porous}\right),$ we have assumed the local thermal equilibrium of fluids and solid materials, which means they have the same temperature at contact point. Also, we take $ \nabla\cdot (\vec{v}T)=\vec{v}\cdot\nabla T $ according to Eq.~$\left(\ref{conservation law 2}\right).$ Eq.~$\left(\ref{darcy's law}\right)$ is the Darcy law, which is valid when both Re and $ k $ are low enough. Owing to $\lambda=-\frac{k}{\eta}$ and $\vec{v}'=\textbf{J}^{\intercal}\vec{v}/(\det\textbf{J}) $, it is easy to get under transformation  $\{x_{i}\}\rightarrow\{y_{j}\}$,
\begin{equation}
v'_{j}=\sum\limits_{i}J_{ji}v_{i}/(\det\textbf{J})=\sum\limits_{ik}J_{ji}\lambda_{ki}\frac{\partial p}{\partial x_{k}}/(\det\textbf{J})=\sum\limits_{ikl}J_{lk}\lambda_{ki}J^{\intercal}_{ij}\frac{\partial p}{\partial y_{l}}/(\det\textbf{J}),
\end{equation}
which means $ \vec{v}'=\lambda'\nabla'p $ and $ \lambda'=\frac{\textbf{J}\lambda\textbf{J}^{\intercal}}{\det\textbf{J}}. $

All the Eqs.~$\left(\ref{heat transfer porous}\right)$, $\left(\ref{darcy's law}\right)$ and $\left(\ref{conservation law 2}\right)$  can keep their form under general coordinate transformation, so we can get the wanted temperature and velocity distribution without changing any property of fluid materials by transforming permeability
\begin{equation}
k'=\frac{\textbf{J}k\textbf{J}^{\intercal}}{\det\textbf{J}},
\end{equation}
thermal conductivity
\begin{equation}
\left\{\begin{array}{ll}\kappa'_{m}=\dfrac{\textbf{J}\kappa_{m}\textbf{J}^{\intercal}}{\det\textbf{J}}\\
\kappa_{f}'=\kappa_{f}\\
\kappa_{s}'=\dfrac{\kappa_{m}'-\phi\kappa_{f}}{1-\phi}
\end{array}\right.
\end{equation}
and the product of density and specific heat at constant pressure
\begin{equation}
\left\{\begin{array}{ll}(\rho C_{p})_{m}'=\dfrac{(\rho C_{p})_{m}}{\det\textbf{J}}\\
(\rho_{f} C_{p,f})'=\rho_{f} C_{p,f}\\
(\rho_{f} C_{p,s})'=\dfrac{(\rho C_{p})_{m}'-\rho_{f} C_{p,f}}{1-\phi}\end{array}\right..
\end{equation}

{\it Thermal cloak and thermal camouflage.}

As a model application of the above theory,
we first attempt to design a thermal cloak in two dimensions. For this purpose, we adopt the geometrical mapping~\cite{S2006} from the original region with radius $r$ satisfying $ 0<r<R_{2} $ to Region~\uppercase\expandafter{\romannumeral2} with radius $r'$ satisfying $R_{1}<r'<R_{2}$:
\begin{equation}
\left\{\begin{array}{ll}r'=R_{1}+\dfrac{R_{2}-R_{1}}{R_{2}}\\ \theta'=\theta\end{array}\right.,
\end{equation}
which is used to cloak an object located in Region I ($0<r<R_{1}$).


Writing all the parameters in Cartesian coordinates, we obtain the Jacobian matrix in Region~\uppercase\expandafter{\romannumeral2}  as
\begin{equation}
\textbf{J}=\left(
\begin{array}{ccc}
\cos\theta & -r'\sin\theta  \\
\sin\theta & r'\cos\theta
\end{array}
\right)
\left(
\begin{array}{ccc}
\frac{R_{2}-R_{1}}{R_{2}} & 0 \\
0 & 1
\end{array}
\right)
\left(
\begin{array}{ccc}
\cos\theta & \sin\theta  \\
-\frac{\sin\theta}{r} & \frac{\cos\theta}{r}
\end{array}
\right), 	
\end{equation}
and it is not difficult to get
 	$ \det\textbf{J}=\frac{R_{2}-R_{1}}{R_{2}}\frac{r'}{r} $ .
So the thermal conductivity in Region \uppercase\expandafter{\romannumeral2} is
$ \kappa'_{m}=\kappa_{m}\frac{\textbf{JJ}^{\intercal}}{\det\textbf{J}} $
and similarly $ k'=k\frac{\textbf{JJ}^{\intercal}}{\det\textbf{J}}$.

 Then we perform finite-element simulations, which combine heat transfer and the Darcy law, by using the commercial software COMSOL Multiphysics ($http://www.comsol.com/$). Here we simulate steady states for simplicity. Fig.~1 shows the basic design for simulations.

We expect to set the background speed as $ v=5\times10^{-3}$  $\rm{m/s} $ with directions along $x$ axis and $y$ axis, respectively. To get the wanted velocity distribution in COMSOL, for example, we can take $ \eta=10^{-3}$  $\rm{Pa\cdot s} ,$ $ k=10^{-12}$  $\rm{m^{2}} $ and $ \Delta p=400$  $\rm{Pa} $ between the two sides of background square when modeling the Darcy law. In this case, Reynolds number Re $ =\frac{\rho R_{2}}{\eta}v=0.1\ll1 $ and $ k\ll (R_{2})^{2} $ so the Darcy law is applicable here.

The simulation results are shown in Fig.~2 and Fig.~3. Firstly we consider the condition that background velocity $ \vec{v} $ is in the $ y $ direction; see Fig.~2. Comparing the first row (cloak) and third row (pure background), we find that the distribution of temperature is  the same in Region~\uppercase\expandafter{\romannumeral3}, so are the distributions of velocity and heat flux. Also, the cloak has a zero temperature gradient in Region~I, so in Fig.~2(c1) the heat flux is zero in Region~\uppercase\expandafter{\romannumeral1}. Actually, we have realized the cloaking of an object in both velocity and temperature fields simultaneously. The conclusion holds the same for the case with a background velocity $ \vec{v} $ in the $ x $ direction, which can be seen in Fig.~3.

Based on the same principle, thermal camouflage, which has been realized in heat conduction~\cite{Qiu}, can also be designed considering thermal convection. In the model of thermal cloak, we add four solid objects in Region \uppercase\expandafter{\romannumeral3} and observe the thermal signals scattered by the four objects. In Fig.~4 and Fig.~5, we show the simulation results for a given pressure difference in different directions. It can be found that the temperature signals (together with the velocity and heat flux signals) in Region \uppercase\expandafter{\romannumeral3} are the same for the cases with and without camouflage devices.

{\it Thermal concentrator.}

A thermal concentrator for heat conduction can be used to enhance the temperature gradient (which means an enlarged heat flux) in a given area. When taking convection into account, we may realize the same effect. To proceed, we consider another geometrical transformation
\begin{equation}
\left\{\begin{array}{ll}r'=\dfrac{R_{1}}{R_{2}}r {\rm \,\,\, as\,\,\,} r<R_{2}, \\
r'=\dfrac{R_{1}-R_{2}}{R_{3}-R_{2}} R_{3}+\dfrac{R_{3}-R_{1}}{R_{3}-R_{2}}r {\rm \,\,\, as\,\,\,} R_{2}<r<R_{3},\end{array}\right.
\end{equation}
which squeezes the region $ 0<r<R_{2} $ to Region I ($ r'<R_{1} $) and then stretches the region $ R_{2}<r<R_{3} $ to Region~II ($ R_{1}<r'<R_{3}$).
 

The Jacobian matrices are
\begin{equation}
\textbf{J}_{1}=\left(
\begin{array}{ccc}
\frac{R_{1}}{R_{2}} & 0  \\
0 & \frac{R_{1}}{R_{2}}
\end{array}
\right)	
\end{equation}
for Region \uppercase\expandafter{\romannumeral1} and
\begin{equation}
\textbf{J}_{2}=\left(
\begin{array}{ccc}
\cos\theta & -r'\sin\theta  \\
\sin\theta & r'\cos\theta
\end{array}
\right)
\left(
\begin{array}{ccc}
\frac{R_{3}-R_{1}}{R_{3}-R_{2}} & 0 \\
0 & 1
\end{array}
\right)
\left(
\begin{array}{ccc}
\cos\theta & \sin\theta  \\
-\frac{\sin\theta}{r} & \frac{\cos\theta}{r}
\end{array}
\right)	
\end{equation}
for Region \uppercase\expandafter{\romannumeral2}. It is easy to derive that

\begin{equation}
\frac{\textbf{J}_{1}\textbf{J}^{\intercal}_{1}}{\det\textbf{J}_{1}}=\left(
\begin{array}{ccc}
1 & 0  \\
0 & 1
\end{array}
\right) .
\end{equation}
So in Region~I ($r'<R_{1}$), we don't need to change $k$ or $\kappa$.

Again we set $ R_{3}=2R_{1}=2\times10^{-5}\ \rm{m}$ and $ R_{2}=1.5\times10^{-5}\ \rm{m}$. All the other parameters are  same with the case of thermal cloak. We simulate two cases with different velocity directions. The simulation results based on the COMSOL Multiphysics are shown in Fig.~2 and Fig.~3. We can see that the heat flux in Region \uppercase\expandafter{\romannumeral1} gathers because of the increased temperature gradient and the amplified velocity in the same region. Meanwhile, in Region \uppercase\expandafter{\romannumeral3} of Fig.~2(a2-c2) and Fig.~3(a2-c2), the temperature, velocity and heat flux are also same as those for the cases of pure background shown in Fig.~2(a3-c3) and Fig.~3(a3-c3).

{\it Discussion and conclusion.}

According to our transformation theory, we can directly write velocity in Region \uppercase\expandafter{\romannumeral2} as
\begin{equation}
\vec{v}'=\frac{R_{2}}{R_{2}-R_{1}}\left(
\begin{array}{ccc}
\frac{r'-R_{1}}{r'}(v_{x}\cos^{2}\theta+v_{y}\sin\theta\cos\theta)+(v_{x}\sin^{2}\theta-v_{y}\sin\theta\cos\theta) \\
\frac{r'-R_{1}}{r'}(v_{x}\sin\theta\cos\theta+v_{y}\sin^{2}\theta)+(-v_{x}\sin\theta\cos\theta+v_{y}\cos^{2}\theta)
\end{array}
\right).
\end{equation}
Here we use $ \vec{v}'=\textbf{J}^{\intercal}\vec{v}/\det\textbf{J} $, not $ \vec{v}'=\textbf{J}^{\intercal}\vec{v} $ .

For concentrator, the theoretical velocity in Region \uppercase\expandafter{\romannumeral2} is
\begin{equation}
\vec{v}'_{2}=\frac{R_{3}-R_{2}}{R_{3}-R_{1}}\left(
\begin{array}{ccc}
\frac{r'+R_{3}}{r'}(v_{x}\cos^{2}\theta+v_{y}\sin\theta\cos\theta)+(v_{x}\sin^{2}\theta-v_{y}\sin\theta\cos\theta) \\
\frac{r'+R_{3}}{r'}(v_{x}\sin\theta\cos\theta+v_{y}\sin^{2}\theta)+(-v_{x}\sin\theta\cos\theta+v_{y}\cos^{2}\theta)
\end{array}
\right)
\end{equation}
and in Region \uppercase\expandafter{\romannumeral3}
\begin{equation}
\vec{v}'_{1}=\frac{R_{2}}{R_{1}}\vec{v}.
\end{equation}
Since $ R_{1}<R_{2} $, the velocity in Region I $ r'<R_{1} $ is amplified indeed. We plot the velocity distributions in Fig.~6 and they agree perfectly with the velocities generated by the Darcy law, as shown in Fig.~2 and Fig.~3.



We have set background speed as $ v=5\times10^{-3}\ \rm{m/s} $ and thus Reynolds number is 0.1. Actually, when considering the thermal cloak or concentrator, the biggest speed is $ v=1\times10^{-2}\ \rm{m/s} $ or $ v=0.75\times10^{-3}\ \rm{m/s} $, meaning that the corresponding Reynolds number is 0.2 or 0.15. This is practical in experiments.



To sum up, we have developed a general transformation theory for manipulating thermal convection, and revealed some novel behaviors of thermal convection, namely, cloaking, concentrating, and camouflage, which bring novel applications in thermal management/manipulation. This work not only   provides new hints on how to control heat transfer by combining thermal conduction, thermal radiation, and thermal convection, but also benefits the research of mass diffusion and other related fields that contain a set of equations and require the transformation of velocities simultaneously.


{\bf Acknowledgments}\\
We acknowledge the financial support by the Science and Technology Commission of Shanghai Municipality under Grant No.~16ZR1445100.


\clearpage
\newpage

\section*{Figure caption}

Fig.~1. Scheme of thermal cloak/concentrator. (a) cloak with a background velocity in the $y$ direction, as indicated by blue lines; (b) cloak with a background velocity in the $x$ direction. (c) concentrator with a background velocity in the $y$ direction; (b) concentrator with a background velocity in the $x$ direction.
For cloaks, Region \uppercase\expandafter{\romannumeral1} ($ r'<R_{1}$) is a circular object to be cloaked while for concentrators, Region \uppercase\expandafter{\romannumeral1}  is the area where heat flux is concentrated.
Region \uppercase\expandafter{\romannumeral2} is the transformation area for cloaks ($ R_{1}<r'<R_{2} $) or concentrators ($ R_{1}<r'<R_{3}$). For concentrators, the white broken-line circle with a radius of $ R_{2} $ means guidelines for the transformation.
In (a) and (b), the blue lines, representing streamlines of flow, would round the object. In (c) and (d), the blue lines would be concentrated in Region~\uppercase\expandafter{\romannumeral1}.
Region~\uppercase\expandafter{\romannumeral3} is the background area and we restrict it within a broken-line square with a side-length of $ 8\times10^{-5}$~$\rm{m} $. We take $ R_{2}=2R_{1}=2\times10^{-5}$~$ \rm{m}$ for cloaks while $ R_{3}=2R_{1}=2\times10^{-5}$~$ \rm{m}$ for concentrators.
For the cloak and concentrator, we put a hot heat source of 303\,K on the left side of the square and a cold heat source of 293 K on the right.
Background material parameters in heat transfer model are $ \rho_{f}=10^{3}$~$\rm{kg/m^{3}},$ $ C_{p,f}=5\times10^{3}$~$\rm{J/(kg\cdot K)}, $
$ \kappa_{f}=1$~$\rm{W/(m\cdot K)}, \rho_{s}=500$~$\rm{kg/m^{3}},$ $ C_{p,s}=500$~$\rm{J/(kg\cdot K)}, $
$ \kappa_{s}=5$~$\rm{W/(m\cdot K)}$, and $ \phi=0.9. $ The object to be cloaked in Region \uppercase\expandafter{\romannumeral1} has the following parameters: $ \rho=10^{4}$~$\rm{kg/m^{3}},$ $C_{p}=5\times10^{3}$~$\rm{J/(kg\cdot K)},$ and
$\kappa=200$~$\rm{W/(m\cdot K)}.$

Fig.~2. Simulation results when the background velocity is along the $y$ direction.
The first row is for the cloak. (a1) describes the temperature distribution with white lines representing isotherms. (b1) shows the speed distribution and the arrows point the direction of velocity, whose lengths are proportional to the speed. Similarly, (c1) shows the distribution of heat flux and the arrows point the direction of heat flow, whose lengths are proportional to the amount of heat flux. The second and third rows represent the cases of concentrator and pure background, respectively. In (a3-c3), the two concentric circles only denote the position for the comparison with (a1-c1) and (a2-c2).

Fig.~3. Same as Fig.~2, but for the background velocity directed in the $x$ direction.


Fig.~4. Simulation results of thermal camouflage for the $y$-directed background velocity. (a1-c1) [or (a2-c2)] respectively describe the distributions of temperature, velocity and heat flux in the existence [or absence] of camouflage device.  The four objects added in Region \uppercase\expandafter{\romannumeral3} have the following parameters: $ \rho=10^{4}$~$\rm{kg/m^{3}},$ $C_{p}=5\times10^{3}$~$\rm{J/(kg\cdot K)},$ and
$\kappa=200$~$\rm{W/(m\cdot K)}.$


Fig.~5. Same as Fig.~4, but for the $x$-directed background velocity.


Fig.~6. Theoretical result of velocity distribution for (a) cloak with background velocity in the $y$ direction, (b) cloak with background velocity in the $x$ direction, (c) concentrator with background velocity in the $y$ direction, and (d) concentrator with background velocity in the $x$ direction. (a,c) [or (b,d)] agree well with the simulation results shown in Fig.~2(b1,b2) [or Fig.~3(b1,b2)].

\clearpage
\newpage

\begin{figure}[!ht]
	\includegraphics[width=1\linewidth]{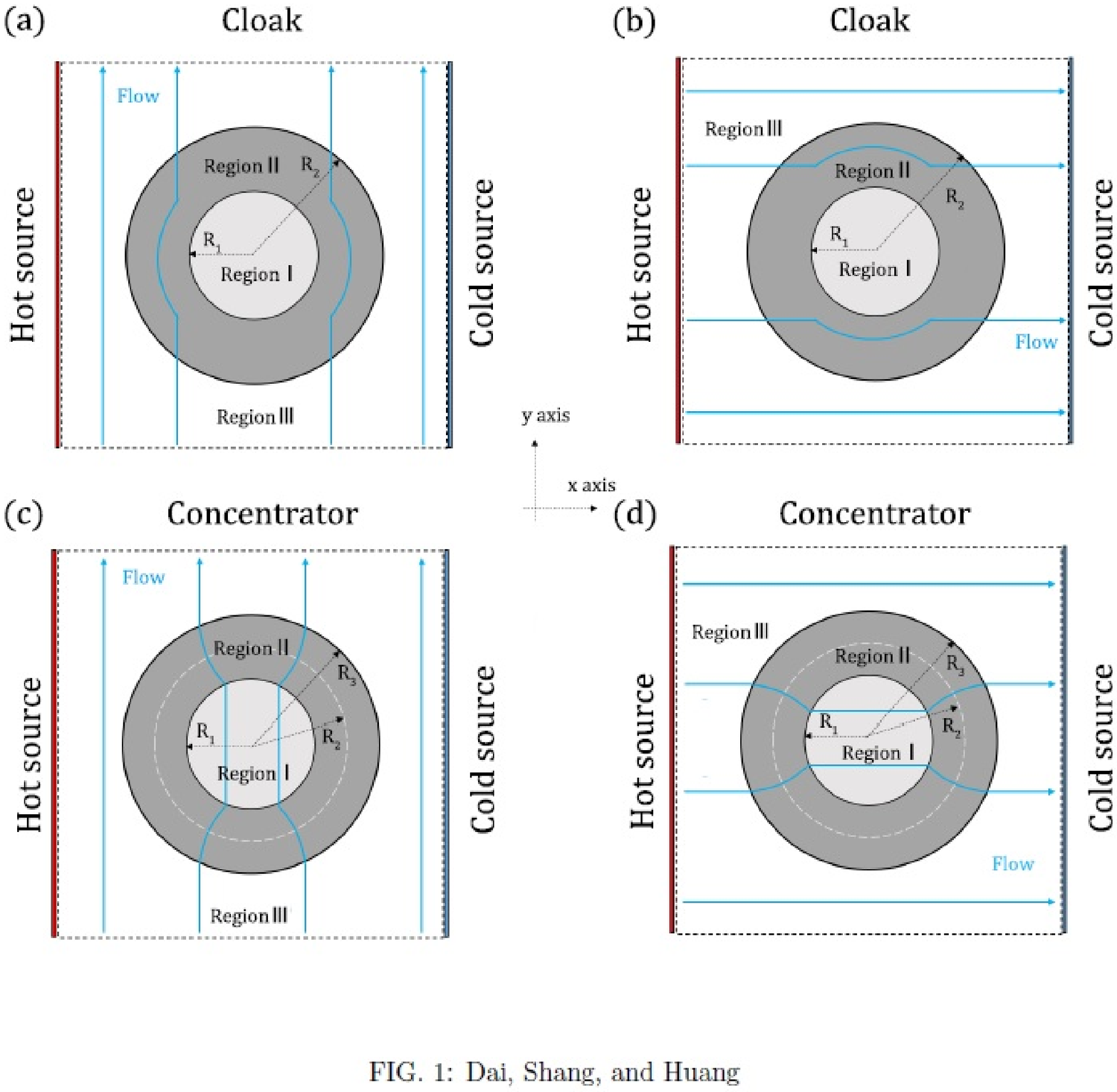}
	\caption{Dai, Shang, and Huang}
\end{figure}

\clearpage
\newpage
\begin{figure}[!ht]
	\includegraphics[width=1\linewidth]{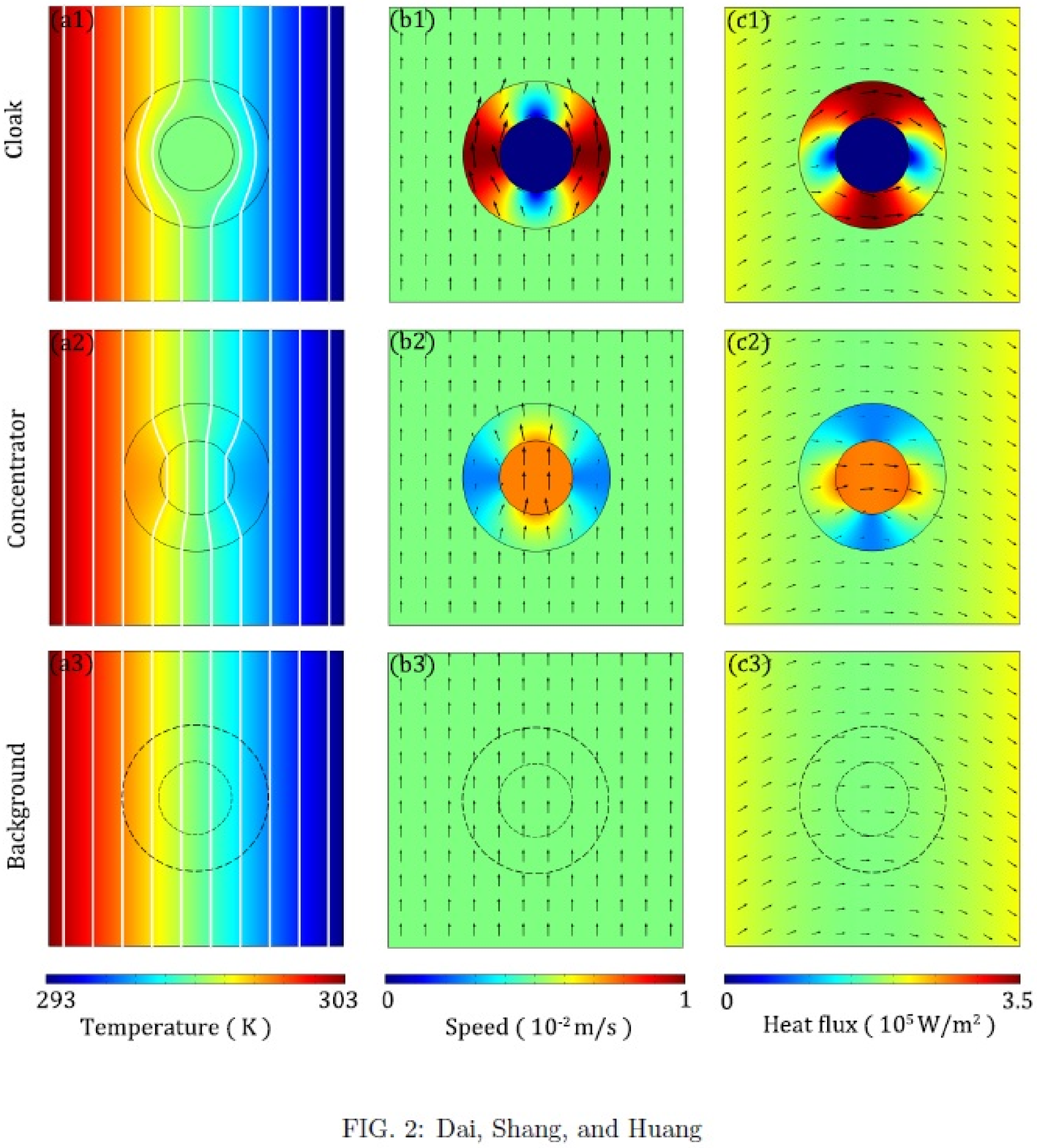}
	\caption{Dai, Shang, and Huang}
\end{figure}

\clearpage
\newpage
\begin{figure}[!ht]
	\includegraphics[width=1\linewidth]{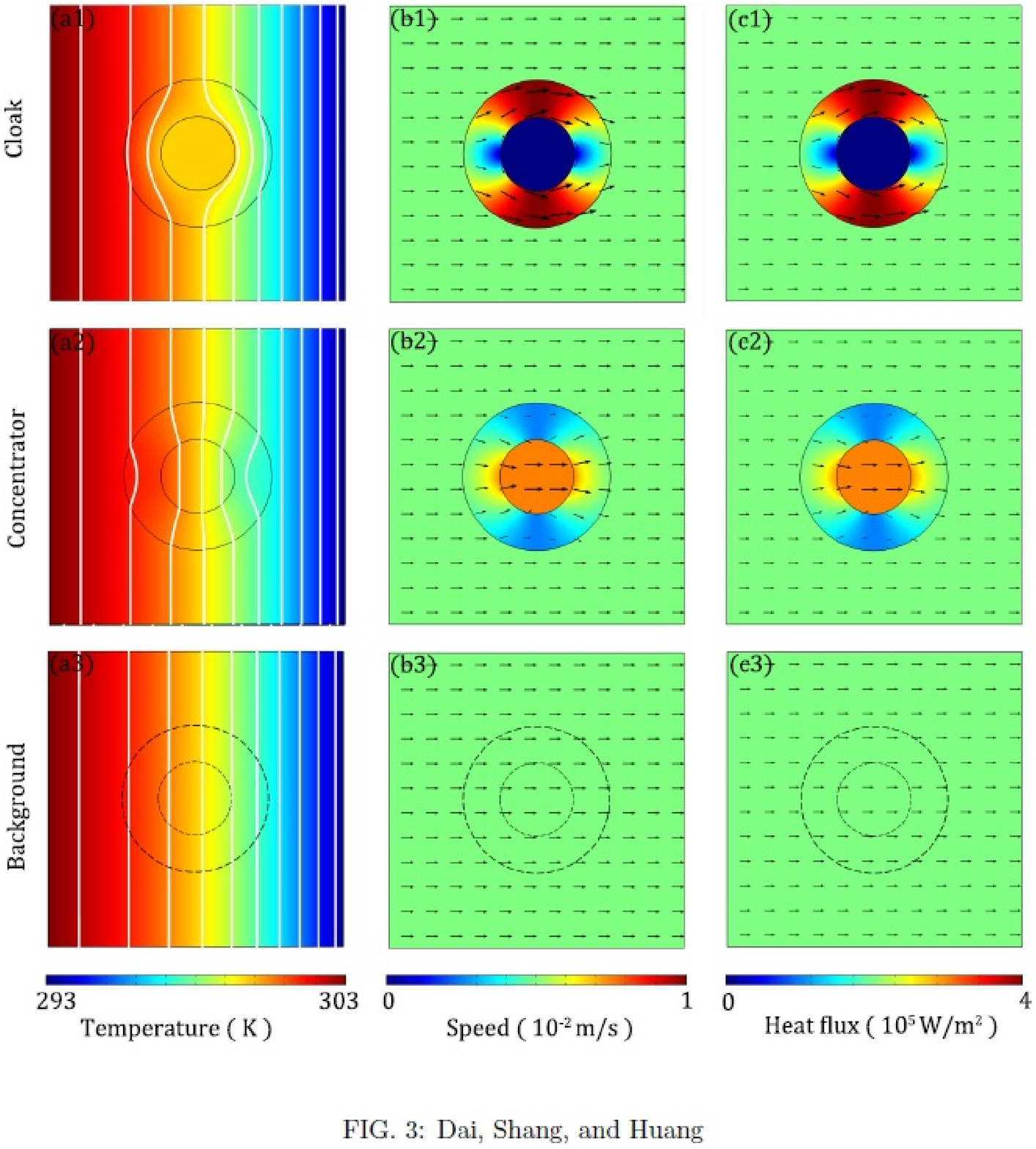}
	\caption{Dai, Shang, and Huang}
\end{figure}

\clearpage
\newpage
\begin{figure}[!ht]
	\includegraphics[width=1\linewidth]{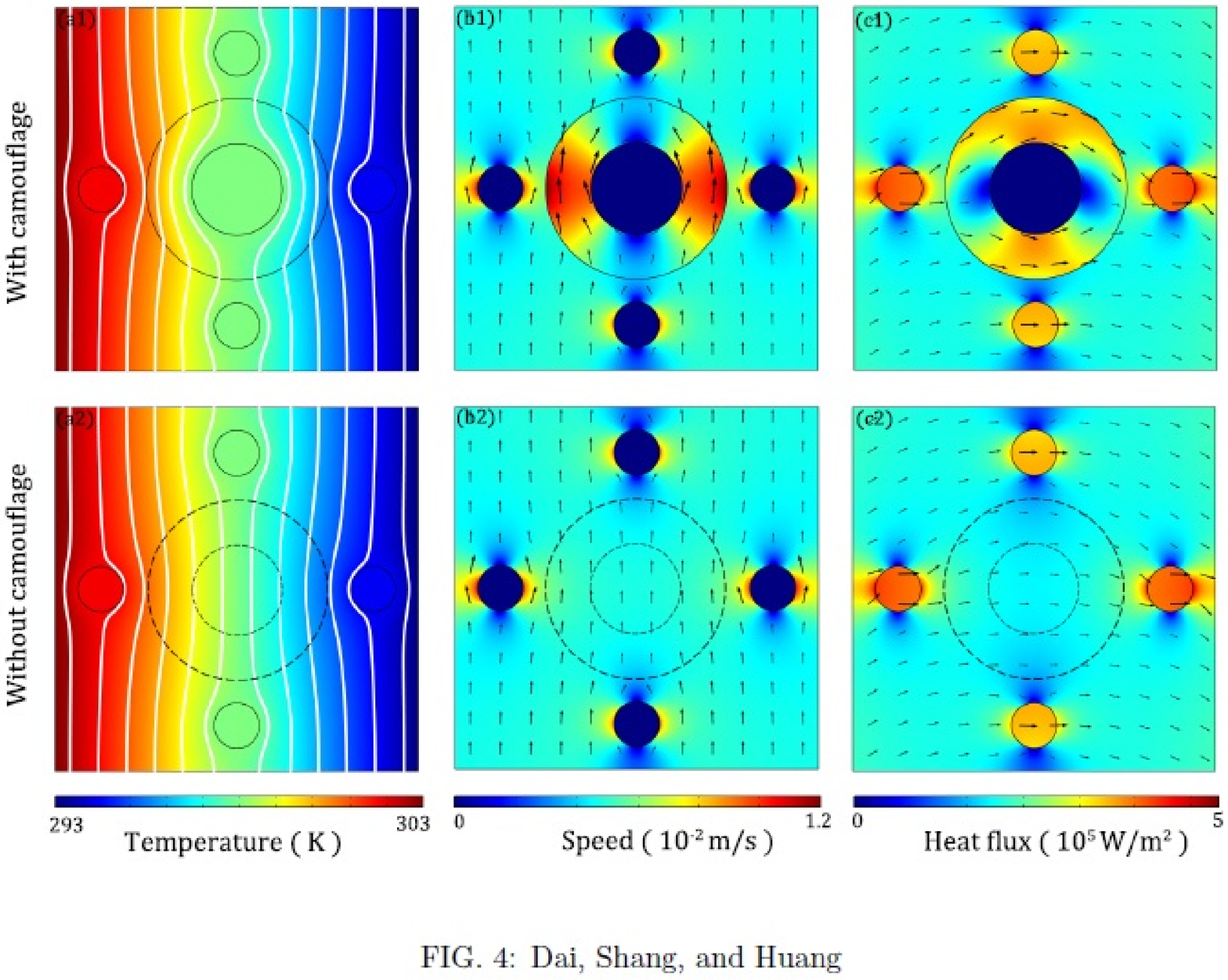}
	\caption{Dai, Shang, and Huang}
\end{figure}

\clearpage
\newpage
\begin{figure}[!ht]
	\includegraphics[width=1\linewidth]{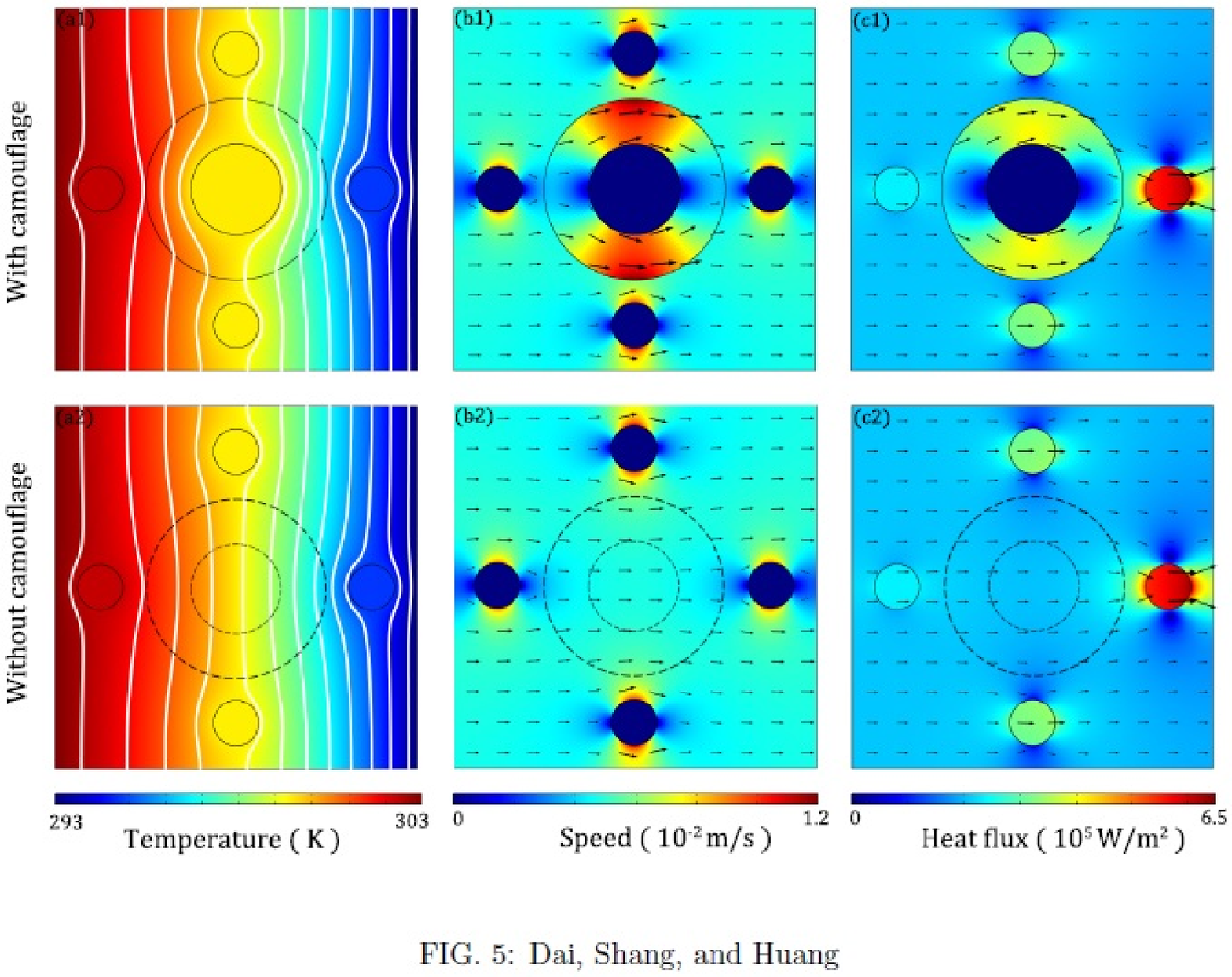}
	\caption{Dai, Shang, and Huang}
\end{figure}

\clearpage
\newpage
\begin{figure}[!ht]
	\includegraphics[width=1\linewidth]{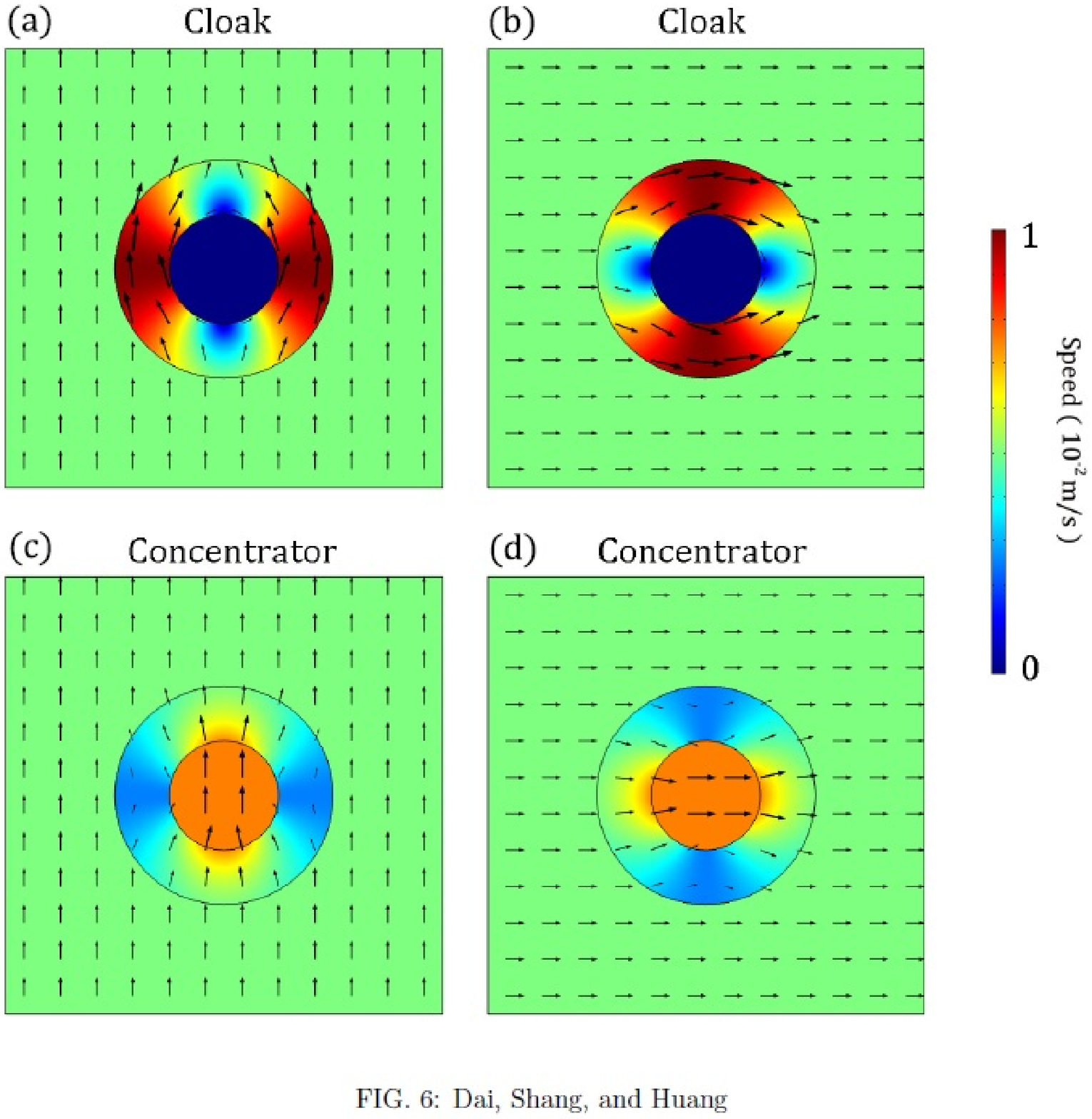}
	\caption{Dai, Shang, and Huang}
\end{figure}

\end{document}